\documentclass[%
 reprint,
 amsmath,amssymb,
 aps,
]{revtex4-2}

\usepackage{graphicx}
\usepackage{dcolumn}
\usepackage{bm}

\usepackage[normalem]{ulem}
\usepackage{color}

\usepackage{caption}

\begin{document}

\preprint{APS/123-QED}

\title{\textbf{Ligament formation of viscoelastic shear- thinning drops impacting on superhydrophobic surfaces}
}
 \thanks{Both authors contributed equally to this work}

\preprint{APS/123-QED}

\title{\textbf{Balloon regime: Drop elasticity leads to complete rebound}
}

\author{Diego Díaz\textsuperscript{1*}}
\author{Arivazhagan G. Balasubramanian\textsuperscript{1,2*}}
\author{Kasra Amini\textsuperscript{1}}
\author{Xiaomei Li\textsuperscript{3}}
\author{Fredrik Lundell\textsuperscript{1}}
\author{Shervin Bagheri\textsuperscript{1}}
\author{Outi Tammisola\textsuperscript{1,2}}
\affiliation{\textsuperscript{\normalfont 1}FLOW, Dept. of Engineering Mechanics, KTH Royal Institute of Technology, 100 44 Stockholm, Sweden}
\affiliation{\textsuperscript{\normalfont 2}Swedish e-Science Research Centre (SeRC), Stockholm, Sweden}
\affiliation{\textsuperscript{\normalfont 3}Institute for Chemical and Bioengineering, Dept. of Chemistry and Applied Bioscience, ETH Zürich, Zürich 8093, Switzerland}

\date{\today}

\begin{abstract}
When a viscoelastic shear-thinning drop of high elasticity hits a superhydrophobic surface, a growing tail-like filament vertically emerges from the impact spot as the contact line recedes. Notably, the ligament transitions into a balloon-like shape before detaching (Balloon regime) completely from the surface. Here, we attribute the ligament formation to the liquid impalement upon impact into the surface protrusion spacing. Our findings reveal that ligament formation can be controlled by tuning the roughness and surface wettability. We show that ligament stretching mainly depends on inertia and gravity, whereas the high elasticity prevents the ligament break up, enabling complete rebounds.

\end{abstract}

\maketitle


The study of drop impact dynamics has fascinated researchers and engineers for over a century due to its fundamental significance in nature~\cite{joung2015aerosol, gilet2015fluid, bourouiba2021fluid} and wide-ranging applications such as surface printing methods~\cite{modak2020drop, lohse2022fundamental}, energy harvesting~\cite{PhysRevLett.125.078301}, heat transfer~\cite{shiri2017heat}, and beyond~\cite{yarin2006drop, mohammad2023physics}. Although the phenomenon is critical in many industrial processes, it can often lead to undesired effects such as splashing~\cite{yarin2006drop} and surface damage~\cite{fujisawa2012experiments, poloprudsky2021surface, nastic2023high}. In particular, drop impact on superhydrophobic surfaces has garnered significant attention due to its potential for complete droplet rebound—a key feature for self-cleaning~\cite{milionis2019engineering}, anti-icing~\cite{zhu2020transparent}, and drag reduction purposes~\cite{zhang2021review}. While the dynamics of Newtonian drops in these scenarios are relatively well understood, the behavior of non-Newtonian viscoelastic drops, especially at high impact speeds, remains poorly explored. This gap in understanding also limits our ability to design surfaces that can effectively repel complex liquids under high impact forces while mitigating surface damage.

Viscoelasticity, imparted by the addition of polymers to water, introduces unique features during drop impact, such as the formation of elongated threads that interact with surface microstructures~\cite{xu2016sliding, chen2018impact}. These interactions challenge the conventional non-wetting behavior of superhydrophobic surfaces and raise fundamental questions about how elasticity and shear-thinning influence droplet dynamics. Importantly, achieving simultaneous droplet rebound and suppression of breakup/splashing at very high impact speeds has proven elusive~\cite{bergeron2000controlling, bartolo2007dynamics, chen2018impact, dhar2019onset}. Thereby, a critical and unresolved question is whether polymer additives can facilitate droplet rebound without splashing at very high impact speeds—-conditions where Newtonian droplets typically eject secondary drops from the spreading rim, which results in a splash.

In this Letter, we show that polymer additives can indeed restore droplet rebound on superhydrophobic surfaces at very high impact speeds, a feat previously not observed for non-Newtonian drops. Through experiments and numerical simulations, we discover a novel regime—termed the Balloon regime—characterized by the formation and complete detachment of a vertical ligament during droplet rebound. This ligament-driven rebound is governed by the interplay of elastic stresses, inertia, and liquid penetration into surface microstructures. Our findings introduce a new approach for controlling droplet dynamics by leveraging the rheology of polymer-infused liquids and tailored preparation of surface microstructures. This provides a pathway to design conditions that retain liquid-repellent properties at high impact speeds, without leaving satellite drops. Here, we will outline the key physical factors responsible for \emph{Balloon regime} by experiments and numerical simulations.

\begin{figure*}
    \includegraphics[width=0.8\textwidth]{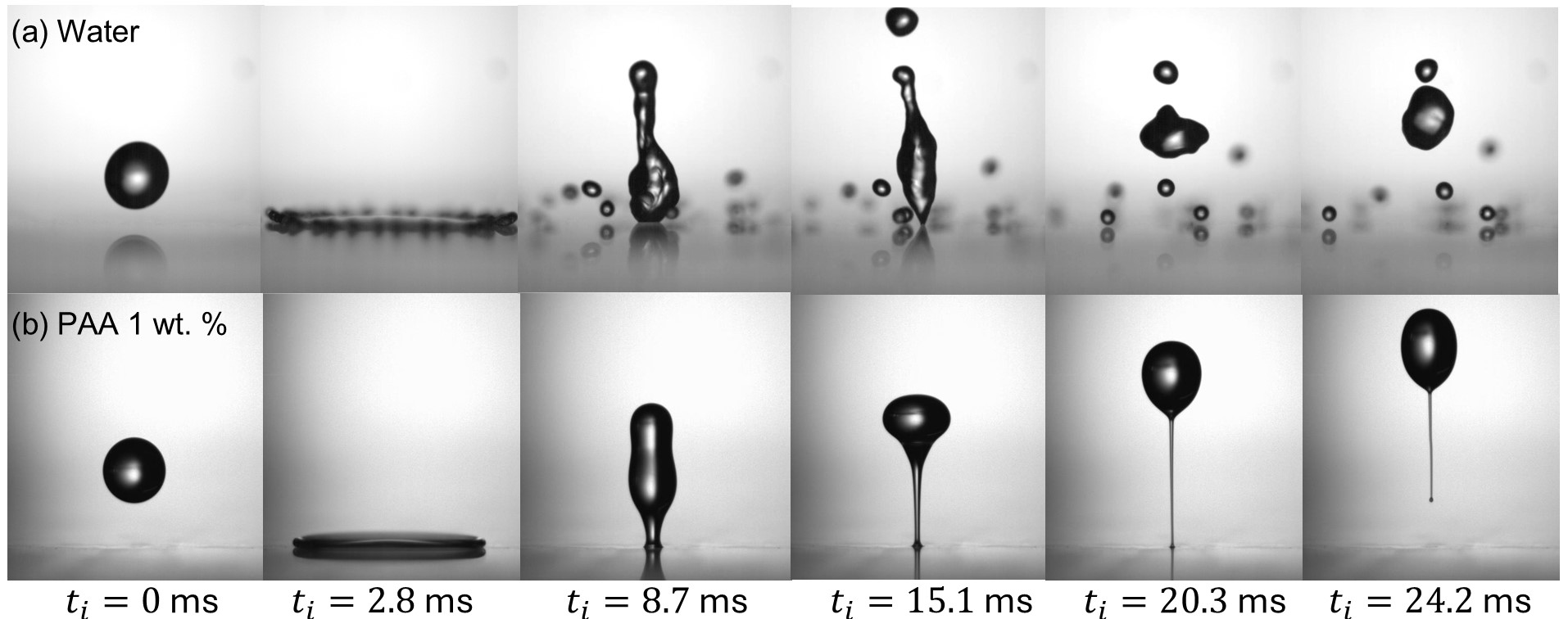}
    \caption{\label{fig:1} $(a)$~Time-lapsed snapshots of a water drop and $(b)$~PAA 1\%wt drop impacting the super-hydrophobic surface at $We =208$. }
\end{figure*}
Viscoelastic aqueous drops of initial diameter $D_0 =2.5$~mm and impact speed $V_0$ ranging from $0.23$ to $3.4$~m/s were dispensed using a stainless needle attached to a syringe pump onto a super-hydrophobic Glaco surface. The surface was prepared by spray-coating three times glass slides with silica nanoparticles after plasma activation (see Supplemental Material, Sec. I~\cite{supplement} for experimental details). All PAA concentrations practically showed the same static contact angles as water ($\Theta_s \sim 167^{\circ}\pm~2$). The viscoelastic liquid solution is prepared by mixing Polyacrylamide (PAA, Mw $>$ 15$\times$ 10$^{6}$ Da) in deionized water in concentrations $C_w$~(by mass) between 0.025 and 1~wt.~$\%$. Rheometric measurements evidenced shear-thinning for all solutions, with zero shear viscosities ranging from $7$ to $25 \times 10^{4}$ mPa s, and elastic relaxation times between $0.6$~ms to $7$~s. 
The impact process was recorded simultaneously from the side and bottom views by two high-speed cameras at up to 4900 fps.
 
The Weber number $We = \rho v_0^{2} D_{0}/ \sigma$ represents the variation of drop inertia over capillary pressure ($2<We<408$), where $\rho$ is fluid density and~$\sigma$ denotes the surface tension coefficient. When $2<We<136$, drops rebound completely for both water and viscoelastic liquids (See Supplemental Material, Sec. II~\cite{supplement}). Only for water, secondary drops are ejected from the top part of the primary drop. When $We > 136$, water drops exhibited splashing behavior~(Fig.~\ref{fig:1}a), with complete surface detachment, but forming numerous satellite drops behind. However, splashing is completely suppressed for viscoelastic drops, which during the receding phase ($C_m >$0.025 $\%$) form a ligament that extends and becomes thinner over time, whereas a head droplet arises at the top (Fig.~\ref{fig:1}b, see Supplemental Movie S1~\cite{supplement}). The ligament is more prominent at higher polymer concentrations. Eventually, the ligament as a whole detaches from the surface giving rise to a spike at the bottom (tail) that retracts back to the primary drop. We can remark on two important aspects of this phenomenon. First, a complete rebound is achieved for the entire range of $We$, contrarily with previous studies where bouncing behavior can be suppressed by increasing polymer concentration. Second, just right before bouncing, the head drop gives rise to a balloon-like shape for 0.5\% and 1\% wt. Such behavior has not been reported before for impacting drops and evidences the rich dynamics of viscoelastic/non-Newtonian droplet impact. 

We show that the ligament formation and resulting balloon instability/filamentous rebound is intimately linked with the Cassie-Baxter~\cite{Cassie1944} to Wenzel ~\cite{wenzel1936resistance} transition on the superhydrophobic surface. The Glaco surface is formed by self-assembled silica nanoparticles with a fractal type structure. This allows the formation of air pockets, reducing the surface energy and increasing the repellency to water. When a drop hits a structured surface, the balance between the wetting and antiwetting pressure determines the wetting states~\cite{bartolo2006bouncing,reyssat2006bouncing}. Wetting pressure is given mainly by the Hammer pressure $P_H=\rho C v_0$/5~\cite{deng2009nonwetting}, with $C$ the speed of sound in water (1497 m/s \cite{del1972speed}), while the anti-wetting pressure is the capillary pressure $P_C=-2\sqrt{2} \sigma \cos{\theta_a}/r$~\cite{deng2009nonwetting,zhao2017impact}, where $\theta_a$ is the advancing contact angle of the flat surface($\approx 120$°) and $r$ is the spacing between the surface microstructures. When $P_H=P_C$, $r\sim 5 \sigma/\rho C V_0$, which at $We=136$ (where we start to observe ligament formation) is $\sim 0.1~ \mu$m. This is in good agreement with the topography of the Glaco surface~\cite{langley2018air}, suggesting that liquid penetrates into the spacing between the micro-structures directly at impact. In fact, our numerical simulations will show in the next figures a maximum peak of pressure at a moment close to the contact time with the surface.

\begin{figure}[htb!]
    \includegraphics[width=\columnwidth]{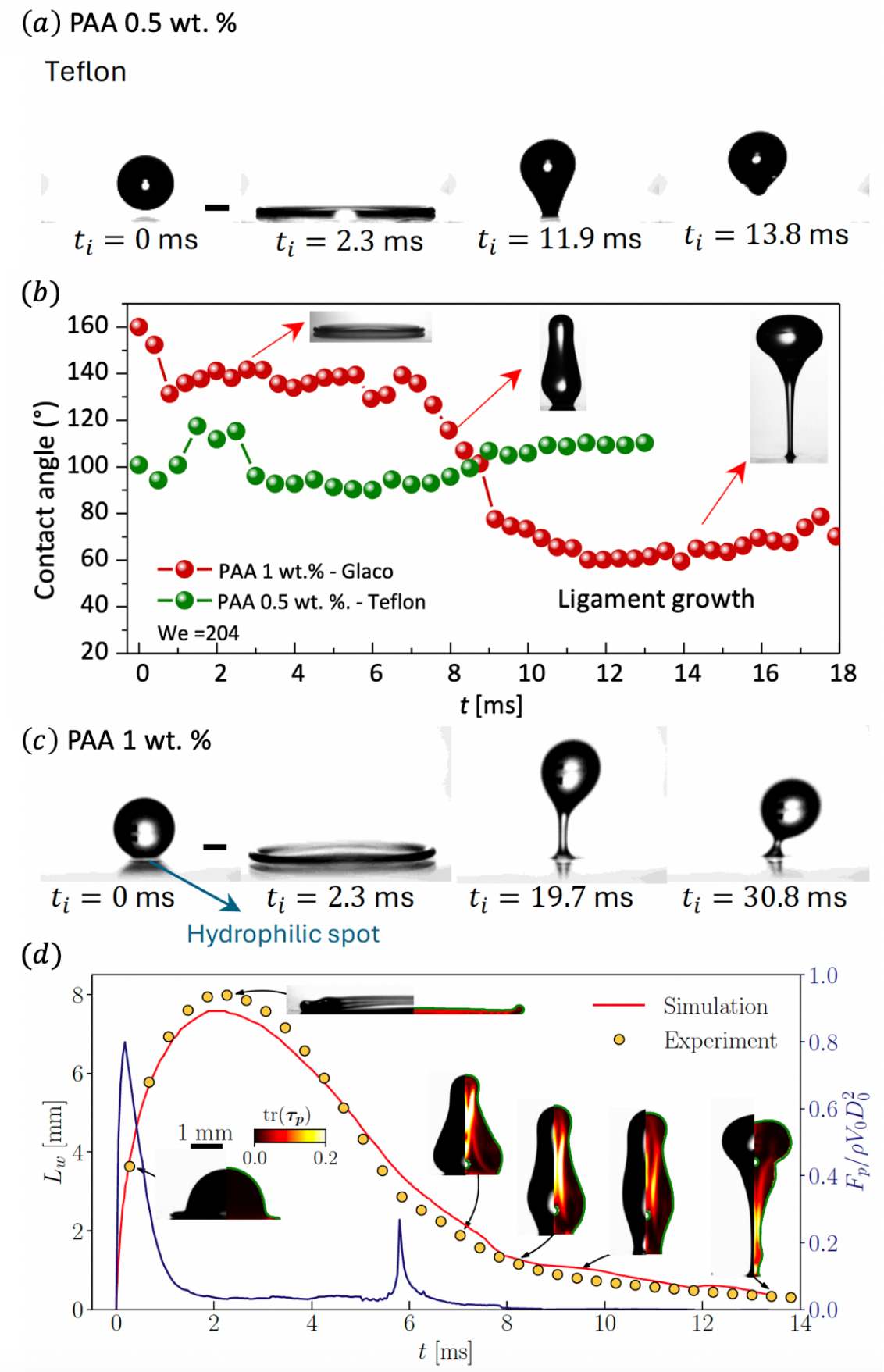}
    \caption{\label{fig:2} $(a)$~ PAA 0.5~wt.~$\%$~droplet impacting a hydrophobic Teflon surface.~$(b)$~ Dynamic contact angle over time for the case shown in $(a)$(green dots, rebound without ligaments) and in Fig.~\ref{fig:1}b) (red dots, rebound with ligaments).~$(c)$~  PAA 1~$\%$wt droplet impacting on Glaco surface with a hydrophilic spot. Scale bar represents $1$~mm.~$(d)$~Temporal evolution of the wetting length (left axis) captured by experiments and numerical simulation for PAA 1 wt.~\% and $We = 272$, with the corresponding non-dimensional wall pressure force plotted on the right axis. The insets correspond to snapshots at different time instants with experimental data plotted on the left and the non-dimensional trace of polymer stress, obtained from simulations plotted on the right.}
\end{figure}

To verify the role of Hammer pressure and Cassie-Wenzel transition, we performed droplet impact experiments on a smooth hydrophobic Teflon AF film layer ($\sim$~60 nm, $\Theta_s=120\pm 2$) on a sputter-coated glass slide. Preparation methods were according to~\cite{li2022spontaneous}. We compared PAA concentrations and $We$ values where ligaments were observed on Glaco surface. Notably, ligament formation was completely suppressed by the hydrophobic surface (Fig.~\ref{fig:2}a and Fig.~S4, Supplemental Material, Sec. III~\cite{supplement}). The lower roughness of Teflon (rms $\sim 5$ nm) evidences its smoothness and therefore the absence of micro-structures with air pockets. As a result, liquid impalement is impossible, which allows the droplet to remain in the Wenzel state during the whole impact process. Thus, the emergence of ligaments is clearly a surface-dependent phenomenon.

An additional indicator of the Cassie-Wenzel transition is the variation of the dynamic contact angle throughout the entire impact process~(Fig.~\ref{fig:2}b). We measured the dynamic contact angle by a tangent fitting method according to~\cite{shumaly2023deep}. For a complete rebound without ligaments, the contact angle during the drop recoil is constant~(Fig.~\ref{fig:2}b, green markers). For the case with ligament formation, the dynamic contact angle remains first at $\sim140$° during spreading and receding, but decreases abruptly when a ligament starts to form, reaching a plateau at $\sim60$°. Liquid impalement during spreading implies the removal of the air pockets. Consequently, when the contact line recedes over the wetted area, the liquid will directly be in contact with the silica nanoparticles. This increases the adhesion, reducing significantly the contact angle and the contact line velocity. To mimic such change in the adhesion, we prepared a Glaco surface with a hydrophilic spot of $0.8$~mm diameter.  Subsequently, we dispensed PAA drops on this surface with the same conditions as Fig.~\ref{fig:1}b. The experiment revealed the formation of a shorter ligament pinned to the spot that grows without detachment (Fig.~\ref{fig:2}c). Thus, an abrupt increase in wettability is a key requirement to generate ligaments. 

We imposed the dynamic contact angles from experiments to our interface-resolved numerical simulations~(Fig.~\ref{fig:4}, See Supplemental Material, Sec. IV~\cite{supplement} for details on the numerical setup) for a qualitative comparison. The wetting length, $L_w$ and interface profiles display a good qualitative agreement for PAA 1 wt. \% at $We = 272$. The non-dimensional wall pressure force over time indicates a primary pressure peak at the impact much higher than the secondary peak at Worthington jet formation~(see Fig.~\ref{fig:4}), similar to the observation of drop impact with Newtonian fluids~\cite{zhang2022impact}. This means that impalement should occur during impact rather than the receding phase. The trace of polymer stress, which is a measure of deformation or stretching of polymer chains is plotted at different time instants in the inset of Fig.~\ref{fig:4}. During the initial stages of drop impact, viscous effects dominate over elastic stresses, resulting in a negligible trace of polymer stress. At maximum spreading~($t\approx 2\,\mathrm{ms}$) the polymers are elongated radially. 
\begin{figure}[htb!]
    \includegraphics[angle=-90,width=\columnwidth]{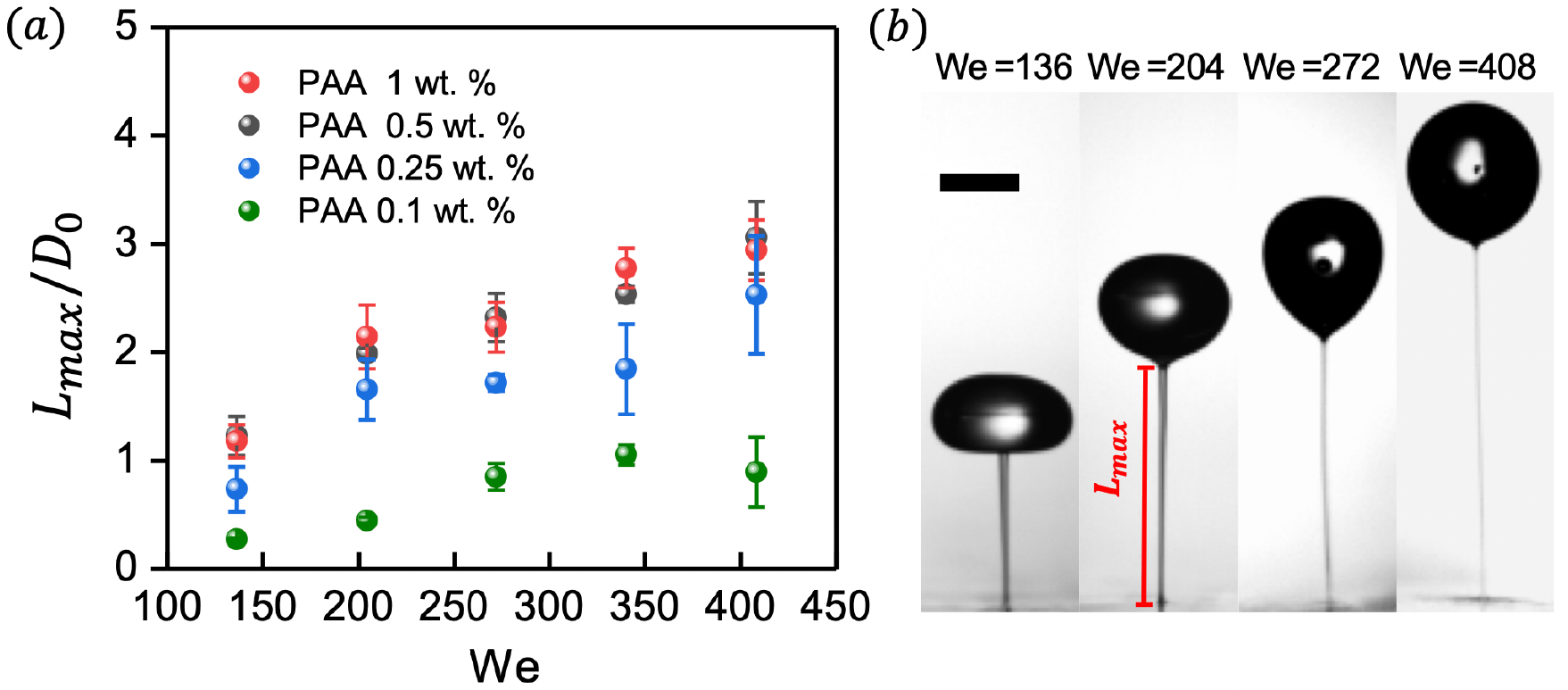}
    \caption{\label{fig:4}~$(a)$~Maximum ligament length $L_{max}$ scaled by the drop diameter as a function of $We$.~$(b)$~Snapshots of $L_{max}$ at different $We$. Scale bar represents $2.5$~mm. }
\end{figure}

At a later stage~$t\approx 6\,\mathrm{ms}$, however, the polymers are highly stretched along the axis near the top, attenuating Worthington jet and supporting the formation of the head droplet. The ligament starts to form after the contact angle changes sign as highlighted in the experiments. For the higher concentration considered in the experiments, the ligament transitions into a shape resembling a balloon rising from the substrate whereby, the liquid flows from the thinning filament to the droplet. We attribute the balloon-shape to the combined role of inertia, elastic and capillary stresses, apart from the minimization of surface energy. At the final stages~($t\approx 13.4\,\mathrm{ms}$), polymers are highly elongated in the thinning filament, preventing its breakup~\cite{bhat2010formation} until it detaches. Complete rebound by ligament detachment is therefore facilitated by the high elasticity of the liquid. The detachment pressure can be estimated as $P_{det}~\sim~\rho V_{ret}^{2}$, where $V_{ret}$ is the retraction velocity of the ligament one frame after detachment. For the case shown in Fig.~\ref{fig:1}b, $P_{det}\sim 19$~kPa, one order of magnitude lower than $P_h$. As a complete rebound for water on Glaco occurs~(refer Fig.~\ref{fig:1}a at~$t_i=15.1\;\mathrm{ms}$) due to the high receding contact angle during recoiling ($\theta_r\sim$~120$^{\circ}$), the detachment of PAA drops with significantly lower $\theta_r$ should be favored by the high elasticity of polymers.

\begin{figure}
    \includegraphics[angle=-90,width=\columnwidth]{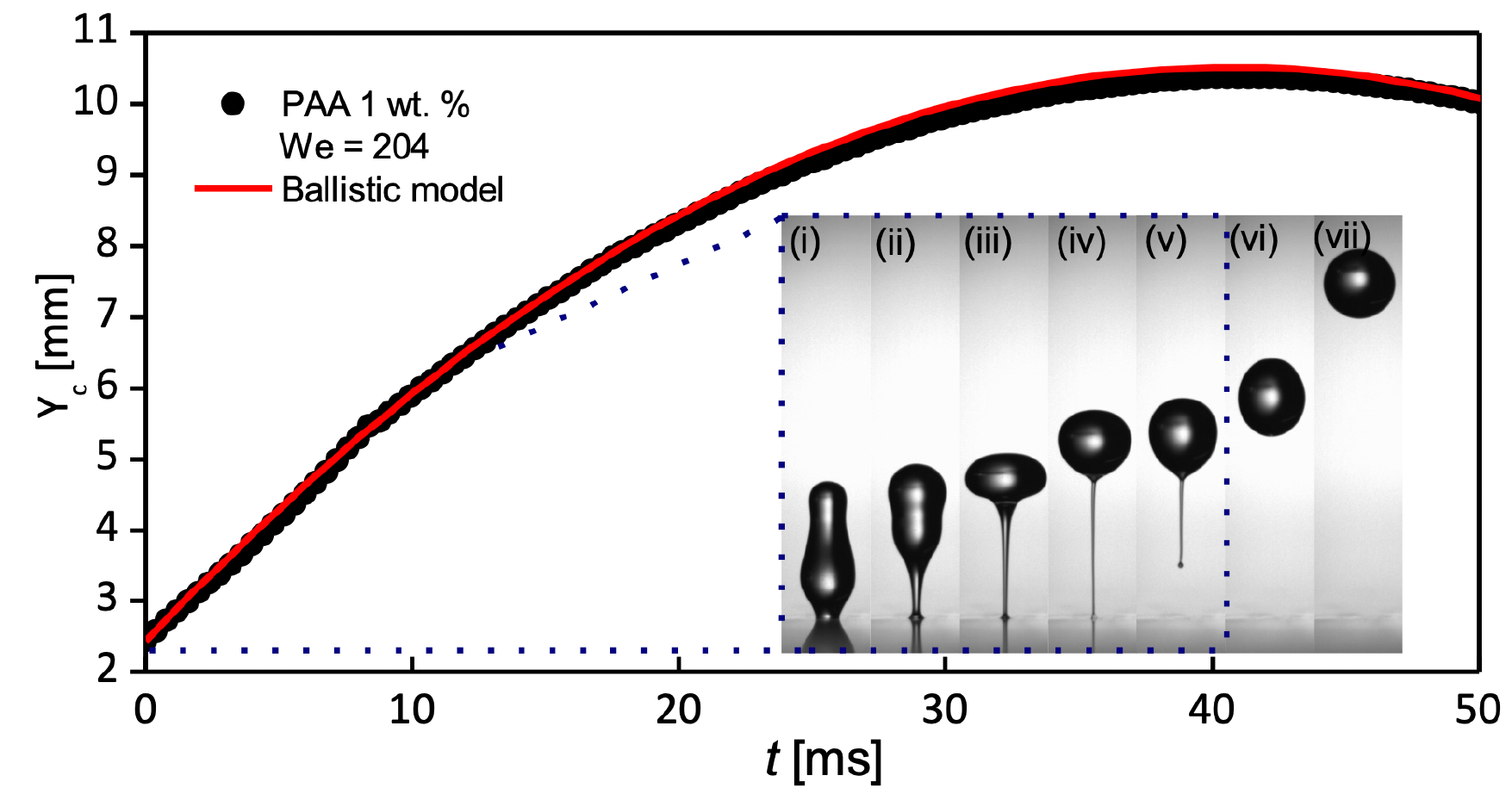}
    \caption{\label{fig:3} Height of the droplet centroid in time for PAA 1 wt.~\%. Insets (i)-(vii) represent the time lapses of the ligament growth: 3, 7, 9, 13, 14, 19, 43 ms, respectively.}
\end{figure}

To elucidate the dominant forces behind the complete drop rebound, the ligament length was measured for all the different concentrations. Fig.~\ref{fig:3}a shows the variation of normalized maximum ligament length $L_{max}$ with respect to $We$ for different polymer concentrations. The plot indicates that $L_{max}$ is proportional with fluid inertia ($L_{\mathrm{max}}\sim We$). The ligament at various $We$ values is shown in Fig.~\ref{fig:3}b, just prior to detachment from the substrate. The increment of $We$ leads to an increase of $P_h$, which causes a deeper liquid impalement. As a result, the time for ligament detachment prolongs while the head droplet moves upwards favoring its growth. This suggests that inertial forces play a key role in the ligament stretching. Indeed, the height of the droplet centroid in time $Y_c(t)$ (Fig.~\ref{fig:3}c,d) can be well described by a ballistic model $Y_c(t)= Y_0+v_{y0}t-\frac{1}{2}gt^{2}$, where $Y_{c_0}$ is the initial vertical position of $Y_c$ when a ligament starts to form (neck emerging below), $v_{y0}$ the vertical speed at that moment ($dY_c/dt$) and $g$ the acceleration of gravity. The integrated difference in potential energy $\Delta E_p$ up to the maximum height between the model and the experiments provides a maximum estimate on the total energy dissipated by viscous and elastic forces. This means that the total dissipative force is $\Delta E_p/L_{max} \sim 10^{-6}$N, which is at least one order of magnitude smaller than the gravitational force $\rho Vg\sim$~8.5~$\times$~$10^{-5}$N (here, $V$ is the drop volume). Therefore, ligament length is determined mainly by the competition between inertial and gravitational forces. Although the elastic stresses are not dominant in determining the ligament length, they are crucial in enabling ligament formation, as explained below.

To conclude, we found a new drop impact regime for viscoelastic droplets on superhydrophobic surfaces, Balloon regime, by ligament formation followed by a complete rebound. We attribute the cause of the phenomenon to the liquid impalement into the surface microstructures (Cassie-Wenzel transition). This is evidenced by a significant decrease in the dynamic contact angle during the receding phase, which is in good agreement with direct numeral simulations. The elastic nature of the liquid allows a stable ligament growth from the impact point until a remarkable complete detachment. The ballistic trajectory of the drop centroid height indicates that ligament length is predominantly controlled by inertia and gravity. We also showed that ligament formation can notably be tuned by surface roughness and wettability. Our findings could be helpful for viscoelastic drop deposition, where the control and prevention of instabilities is crucial in the performance of industrial applications such as inkjet printing and pesticide fabrication. 

The authors acknowledge financial support from the European Union through the European Research Council (ERC) grant no.~``2019-StG-852529, MUCUS", the Swedish Research Council through grant No 2021-04820 and the European Union’s Horizon 2020 research and innovation program under the Marie Skłodowska-Curie grant agreement No. 955605 YIELDGAP. SB gratefully acknowledges support from the ERC project (CoG-101088639 LUBFLOW)” 

\appendix

\nocite{*}
\bibliography{apssamp}

\newpage
\setcounter{equation}{0}
\setcounter{figure}{0}
\setcounter{table}{0}
\setcounter{page}{0}
\makeatletter
\renewcommand{\theequation}{S\arabic{equation}}
\renewcommand{\thefigure}{S\arabic{figure}}
\setcounter{secnumdepth}{5}
\setcounter{enumiv}{42}

\onecolumngrid
\section*{Supplemental Material for Balloon regime: Drop elasticity leads to complete rebound}
\section{Materials and experimental methods}

\subsection{Surface preparation and contact angle measurements}

For the surface preparation, glass slides (1~$\times~$75$~\times$~40~mm$^{3}$) were first Oxygen plasma activated (5 min, 70 W). Afterwards, substrates were sprayed three times with silica nanoparticles suspended in Isopropanol (Glaco Mirror Coat “Zero” from Soft99 Co). Subsequently, a Goniometer Drop Shape Analyzer (DSA25, Krüss) was used to measure contact angles. The static contact angle was measured by the tangent method after depositing a 8~$\mu$L drop Mili-Q water (18~M$\Omega$) on the surface on four different spots.

\subsection{Experimental setup}
Drops are dispensed by a stainless needle attached to a syringe pump (New Era Pump Systems, Inc., USA) at a pump rate of 0.8 $\mu$L/s. The impact process is recorded simultaneously from the side and bottom view by two high-speed cameras (Dantec dynamics, Denmark, 2500-4900 fps). For the bottom view, one mirror is placed surrounding the needle and another below the surface at 45$^{\circ}$ of inclination o deflect the beam from the light source (KL 2500, Schott).

\begin{figure}[h!]
    \includegraphics[width=0.7\textwidth]{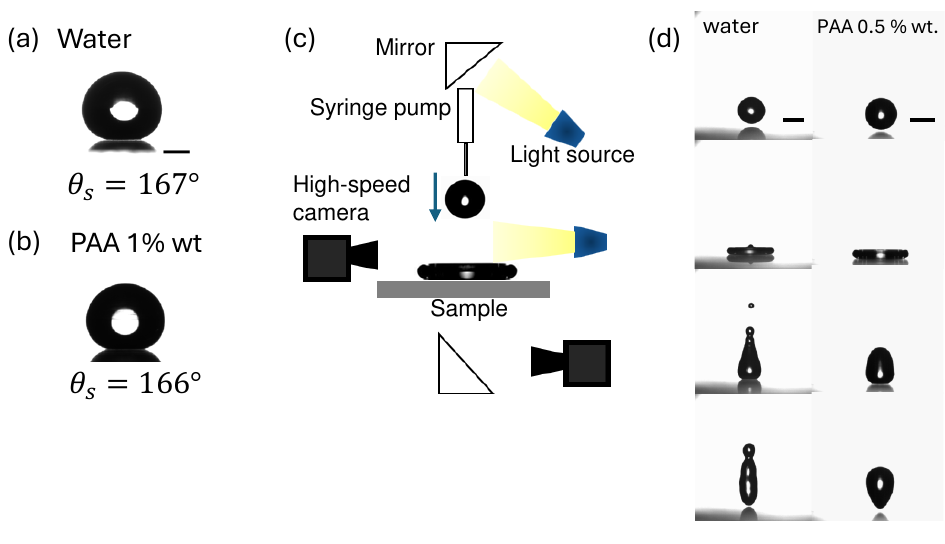}
    \caption{\label{fig:flow} Water $(a)$ and PAA 1\% wt drop $(b)$ deposited on a Glaco superhydrophobic surface.~$(c)$ Schematic of the experimental setup. $(d)$ Water and PAA 0.5 \% wt. impacting at $We=14$ and $We=22$, respectively. Scale bars represents $0.5$~mm in $(a)$ and $2$~mm in $(b)$.}
    \end{figure}
\subsection{Rheology flow curves}~\label{sec:rheology}
Polyacrylamide (PAA) in concentrations of 250, 1000, 2500, 5000, and 10000 ppm (0.025$\%$, 0.1$\%$, 0.25$\%$, 0.5$\%$ and 1$\%$wt.) in deionized water is used. Rheometry of the fluid samples is performed with an Anton Paar MCR 702e Space (Anton Paar, Austria), using a 50 mm cone/plate measuring geometry set (1$^{\circ}$ cone angle, 93 $\mu$m truncation). Steady shear tests in shear rate range $\Dot{\gamma}$ $\epsilon$ [0.001 to 1000] s$^{-1}$, and oscillatory tests were performed, namely strain amplitude sweep in $\gamma$ $\epsilon$ [0.01 to 10 000], and frequency sweep in $w$ $\epsilon$ [0.01 to 300] rad/s. All rheological measurements were performed in the same temperature and environment as the drop impact experiments. The reciprocal of the frequency at the cross-over point of the dynamic moduli ($G'$ and $G''$) in the frequency sweep test is taken as an estimation of the elastic relaxation time. All concentrations show shear-thinning attributes. The extent of shear-thinning increases and zero-shear viscosity plateau moves to lower shear rate ranges with increasing concentration of the polymer solutions (Supplemental Material, S2). From the lowest to highest PAA concentration, zero shear viscosities range from $\sim$ 7 to 25 $\times$ 10$^{3}$ mPa s.

\begin{figure}[h!]
    \includegraphics[width=0.5\textwidth]{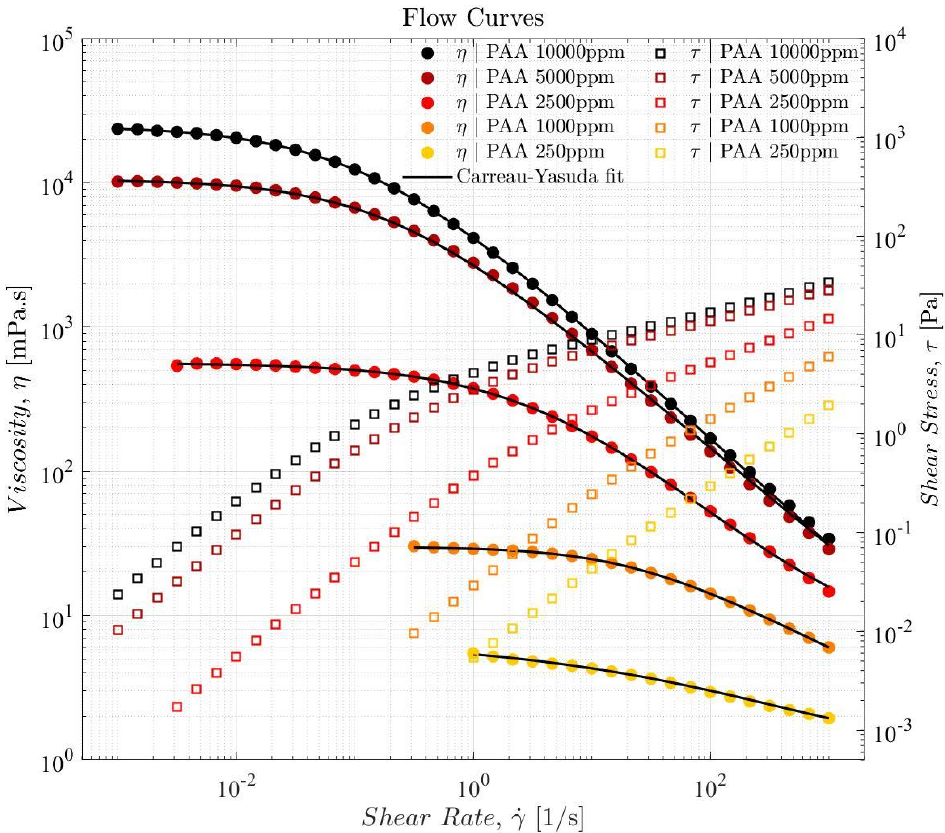}
    \caption{\label{fig:flow} The shear-thinning effects of the polymer solutions are described by the 5-parameter Carreau-Yasuda model, as a function of high/infinite-shear viscosity, $\eta_{inf}$, zero-shear viscosity, $\eta_{0}$, shear thinning relaxation time, $\lambda$, power law index, $n$, and a parameter describing the transition between the power-law region and that of the Newtonian plateau, $a$}
    \end{figure}

    \begin{equation}
    \eta(\dot{\gamma})=\eta_{inf}+(\eta_{0}-\eta_{inf})[1+(\lambda \dot{\gamma})^{a}]^{\frac{n-1}{a}}
\end{equation}
\\

\begin{center}
\begin{tabular}{ |p{2.5cm}||p{1.7cm}|p{1.7cm}|p{1.7cm}||p{1.7cm}||p{1.7cm}| }

 \hline
  Fluid Sample& $\eta_0$ \small[mPa s]  & $\eta_{inf}$\small[mPa s]& $\lambda$\ \small[s] & $n$ & $a$\\
 \hline
 PAA 10000 ppm   & 24810 &1.505  &7.436&   0.2457&0.6509\\
 PAA 5000 ppm&   10570 &1.333 & 4.783   &0.3056&0.7001\\
 PAA 2500 ppm&563.3&6.364& 0.2409&  0.2628&0.6181\\
 PAA 1000 ppm   &29.93 &0.116 & 0.0461&  0.5811&0.8753\\
 PAA 250 ppm&  6.95 &1.156  & 0.0006&0.00003&0.3084\\
 
 \hline

\end{tabular}
\captionof{table}{Carreau-Yasuda fit parameters}
\end{center}

\subsection{Image Processing of dynamic contact angles for droplet impact}

For automated measurements using image processing, we adapted the 4S-SROF toolkit to match our requirements~\cite{shumaly2023deep}. The 4S-SROF toolkit effectively utilizes the OpenCV library~\cite{bradski2000opencv} to manipulate the images, such as separate the drop from its background. We chose to employ morphological transformations for noise reduction, and they proved to be superior to the median filter, guaranteeing the accuracy of our advancing angle measurements~\cite{shumaly2023deep}.  We calculated the advancing angle using the tangent fitting method for the final 10 pixels of the drop near the substrate. The Savitzky-Golay filter~\cite{press1990savitzky} was employed in specific cases to eliminate unwanted noise and enhance the smoothness of the final diagram, facilitating easier interpretation.
\section{Ligament formation at different PAA concentrations}

When a PAA drop of concentrations between 0.1 wt\% to 1\% wt of concentration impacts a Glaco superhydrophobic surface,  a ligament forms as the contact lne recedes. Notably, all the cases show complete ligament detachment and therefore bouncing from the surface.

\begin{figure}[h!]
    \includegraphics[width=\textwidth]{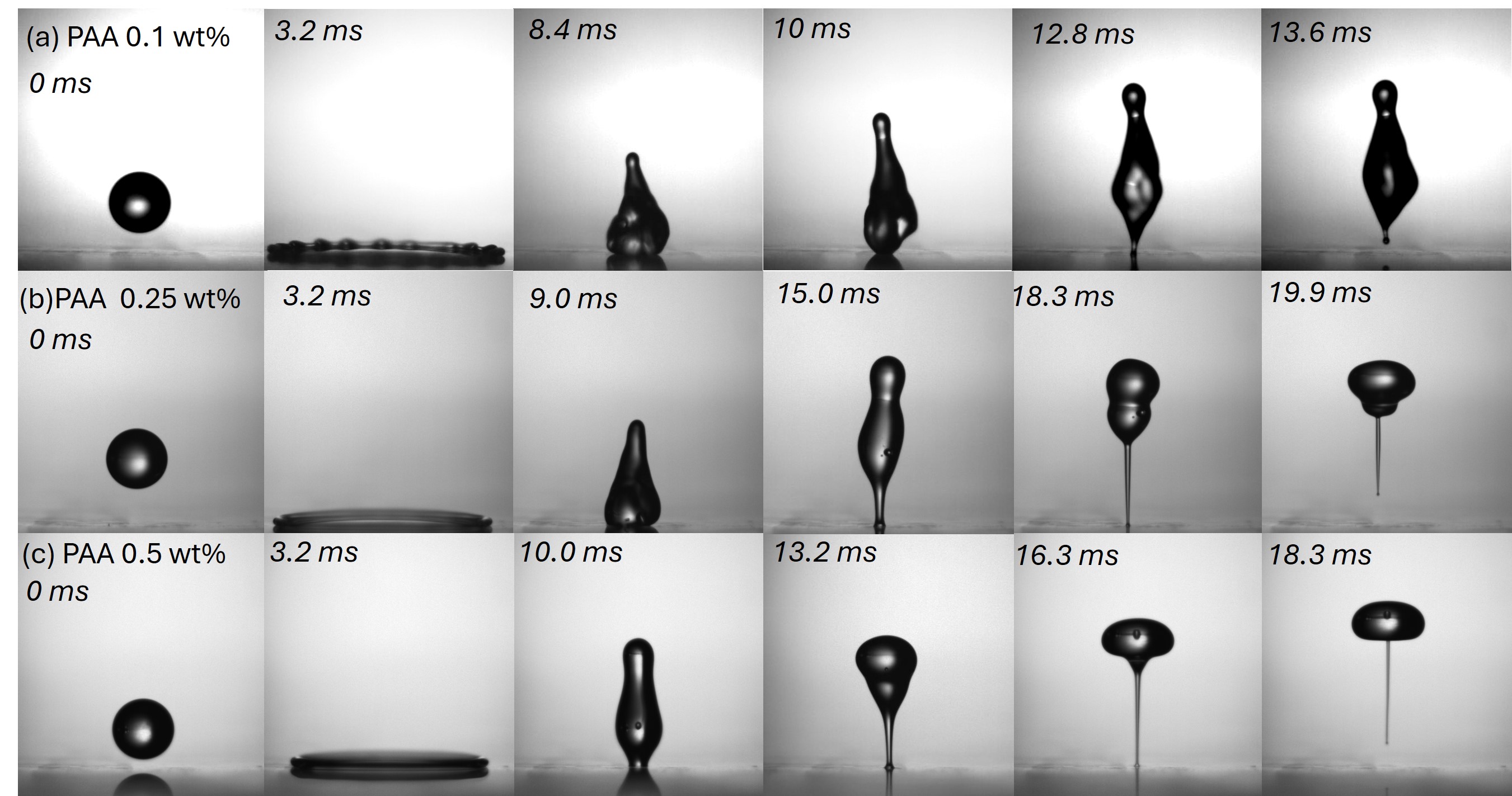}
    \caption{\label{fig:flow} Viscoelastic drop impact on Glaco at different concentrations.}
    \end{figure}

\section{Suppression of ligament formation by a hydrophobic surface}

Drop impact experiments of viscoelastic PAA drops were performed on a smooth hydrophobic surface (Teflon amorphous fluoropolymer (AF), RMS: 5 nm) to prevent liquid impalement. The surface completely suppressed the ligament formation for different concetrations and Weber numbers due to the absence of microstructures with air pockets.

\begin{figure}[h!]
    \includegraphics[width=\textwidth]{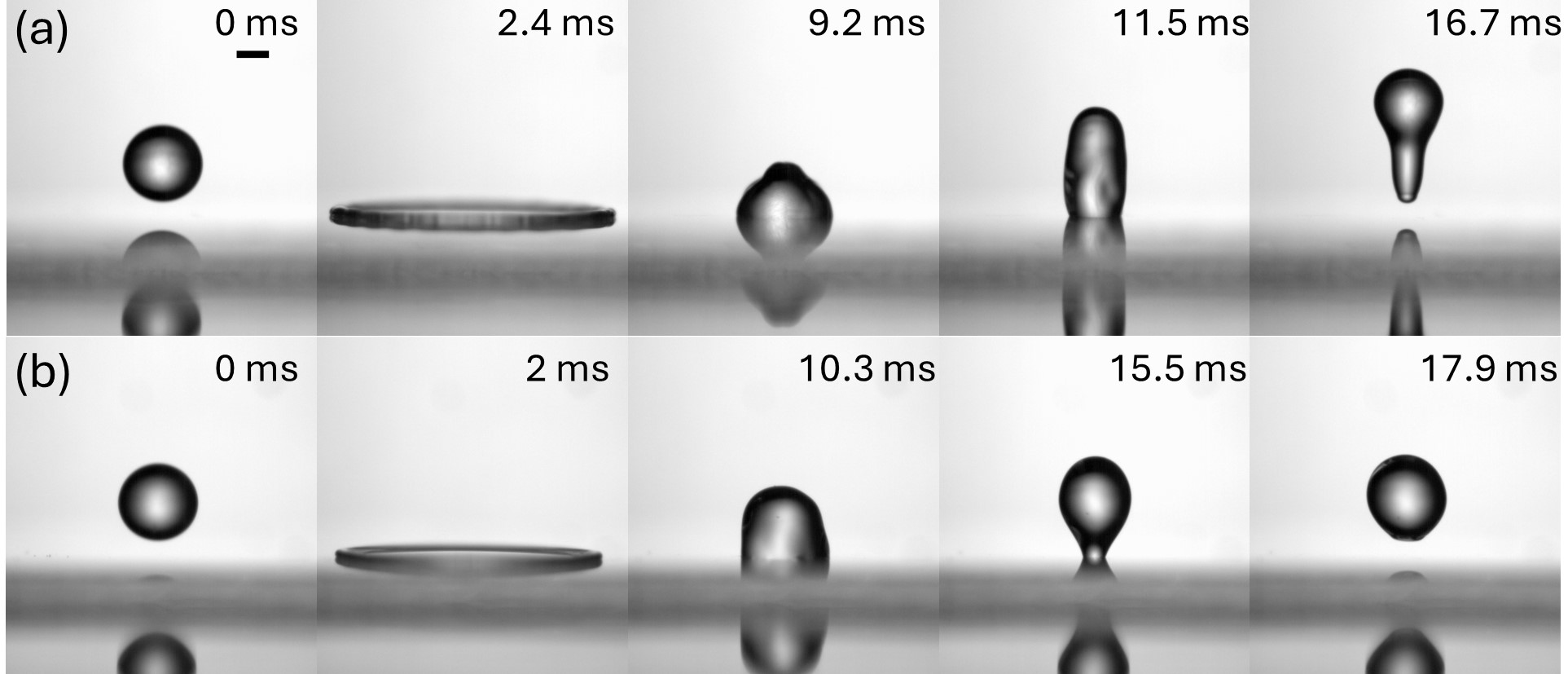}
    \caption{\label{fig:flow} Viscoelastic drop of 2500 ppm~(a)~and 5000 ppm~(b)~impacting onto Teflon AF at $We= 408$) impacting Teflon AF. Scale bar represents $1$~mm.}
    \end{figure}
\section{Ballistic model}
The trajectory of droplet centroid height $Y_c$ is in good agreement with a classic ballistic model: $Y_c(t)= Y_0+v_{y0}t-\frac{1}{2}gt^{2}$, where $Y_{c_0}$ is the initial vertical position of $Y_c$ when a ligament starts to form (neck emerging below), $v_{y0}$ the vertical speed at that moment ($dY_c/dt$) and $g$ the acceleration of gravity. A slight difference between the maximum rebound height between the model and experiments were observed at high Weber numbers, which represents only a dissipated force (viscous and elastic forces) in the order of $\sim10^{-6}$N. This is actually one order of magnitude smaller than the gravitational force acting on the drop ($\sim 8.5\times 10^{-5}$ N).
\begin{figure}[h!]
    \includegraphics[width=\textwidth]{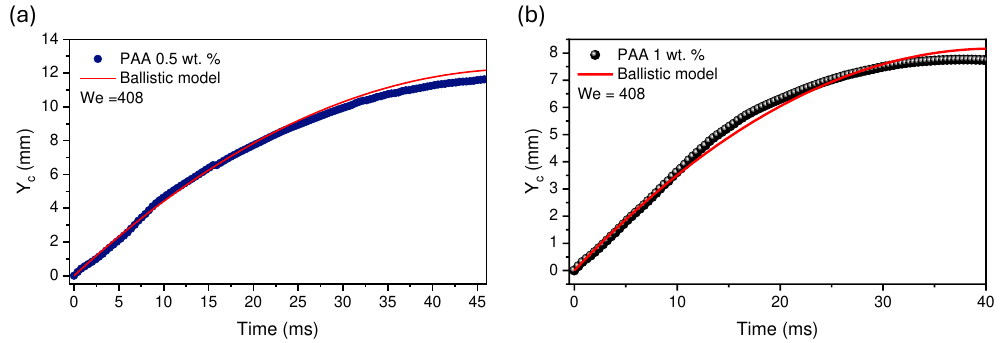}
    \caption{\label{fig:ballistic} Height of the drop centroid over time from the moment when the ligament starts to form until the drop reaches the maximum height after ligament detachment. The plots represents (a)~PAA 0.5 wt \% and (b)~1 wt \% at $We=408$.}
    \end{figure}

\section{Simulation methodology}
\subsection{Constitutive model}
To identify the polymeric stress effects, we perform an axi-symmetric numerical simulation of viscoelastic droplet impinging on a wetting substrate. The interface-resolved simulations are performed using Basilisk C~\citep{S_popinet2009, S_popinet2013}. The volume of each phase of the fluid is tracked with the help of a colour function $c$ ($c = 1$ in liquid and $c = 0$ in gas), which satisfies the scalar-advection equation. The geometrical features of the interface such as its unit vector normal $\boldsymbol{\hat{n}}$ and the curvature $\kappa$ ($= \boldsymbol{\nabla\cdot\hat{n}}$) are calculated using the height-function method \citep{S_popinet2009, S_popinet2018}. We consider a macroscopic continuum description of the fluid-polymer interaction and utilize the FENE-P model or L-PTT model to capture the polymer physics. In the macroscopic description of the homogeneous mixture of polymer dissolved into the solvent fluid, we utilize the second-order statistical correlation of the orientation vector~$\mathbf{R}$ of the spring-dumbbell model, known as the conformation tensor~$\mathbf{A}$ to solve for the polymeric stress. The conformation tensor represents the orientation of the polymer chains and by definition is \emph{symmetric positive definite}.

The dimensional in-compressible Navier--Stokes equation coupled with the evolution equation for polymer conformation tensor is given by,

\begin{align}
     \label{seqn:mom}\frac{\partial \boldsymbol{u}}{\partial t} + \left(\boldsymbol{u} \cdot \nabla \boldsymbol{u}\right) &= -\frac{1}{\rho}\nabla p + \frac{\mu_s}{\rho} \nabla^2 \boldsymbol{u} + \left[ \frac{\mu_p}{\rho \lambda}\nabla \cdot \left( \mathcal{P}(\mathbf{A})\mathbf{A} \right) \right]+\frac{\sigma}{\rho}\kappa\boldsymbol{n}\delta_s\,,\\
     \nabla \cdot \boldsymbol{u} &= 0\,,\\
     \label{seqn:peqn}\frac{\partial \mathbf{A}}{\partial t} + \boldsymbol{u}\cdot \nabla \mathbf{A} &= \mathbf{A}\cdot \nabla \boldsymbol{u} + \left(\nabla \boldsymbol{u}\right)^T\cdot \mathbf{A} - \frac{1}{\lambda} \mathcal{P}(\mathbf{A})\,,
\end{align}
where~$\boldsymbol{u}$ is the velocity with corresponding components in radial and wall-normal directions represented by~$u_r,u_z$ and~$p$ is the pressure with~$t$ denoting the time.  In equation~(\ref{seqn:mom}),~$\rho$ is the density of the fluid and~$\lambda$~denotes the relaxation time-scale of polymer stress. The solvent and polymer viscosity of the liquid droplet is given by~$\mu_s,\,\mu_p$, respectively and it depends on the polymer concentration. The solvent-to-total viscosity ratio is defined as~$\beta := \mu_s/(\mu_s + \mu_p)$. In the one fluid formulation, the fluid properties vary as $\rho(c) = c\rho_1 + (1-c)\rho_2; \mu(c) = c\mu_1 + (1-c)\mu_2$, where $\rho_1, \rho_2$~($\mu_1,\mu_2$) are the density~(viscosity) of liquid and air, respectively. The surface tension coefficient is denoted by~$\sigma$ with the unit vector normal to the interface given by~$\boldsymbol{n}$ and~$\delta_s$ denotes the surface Dirac-delta function in the continuum surface force model~\cite{S_brackbill1992}. The relaxation function~$\mathcal{P}(\mathbf{A})$ differs for the considered viscoelastic model with,
\begin{equation}
     \mathcal{P}(\mathbf{A}) = \frac{L_\mathrm{max}^2}{L_\mathrm{max}^2 - \mathrm{tr}(\mathbf{A})}\,,
\end{equation}
for the FENE-P model. Here,~$L_\mathrm{max}^2$ accounts for the finite length of the polymer molecules and denotes the upper limit of the normalized polymer extension length where, the polymers cannot absorb more energy from the flow. In the present simulations,~$L_\mathrm{max}^2$ is set to 3600. For the L-PTT model, the relaxation function is defined as 
\begin{equation}
    \mathcal{P}(\mathbf{A}) = 1 + \epsilon\, \mathrm{tr}(\mathbf{A}-\mathbf{I})(\mathbf{A}-\mathbf{I})\,,
\end{equation}
with $\epsilon$ indicating the extensibility parameter, which is set to $0.14$ in the present investigation.

The polymer stress~$\tau_p$ can be retrieved from the conformation tensor using Kramer relationship,
\begin{equation}
    \boldsymbol{\tau_p} = \frac{\mu_p}{\lambda} \mathcal{S}(\mathbf{A})\,,
\end{equation}
with $\mathcal{S}$ indicating the recoverable strain function. $\mathcal{S}(\mathbf{A}) = \mathbf{A} - \mathbf{I}$ for the L-PTT model and it is the same as~$\mathcal{P}(\mathbf{A})$ for the FENE-P model~\cite{S_comminal}. In the present study, log-conformation approach~\cite{S_fattal2004} is utilized to overcome the high Weissenberg number problem.

\subsection{Viscoelastic liquid rheology}
Our aim is to replicate the experimental observation of drop impact of PAA solution and hence we consider L-PTT constitutive relationship~(based on network theory for polymer dynamics) to adequate model the elastic characteristic of the material. Furthermore, the motivation for using the L-PTT model lies in its ability to tune the extensibility parameter, which balances the shear-thinning and extensional hardening behavior of the viscoelastic material. This is particularly important for capturing ligament formation, where the flow closely resembles a~\emph{uni-axial extensional flow}.

\begin{figure}[hbt]
    \includegraphics[width=0.5\columnwidth]{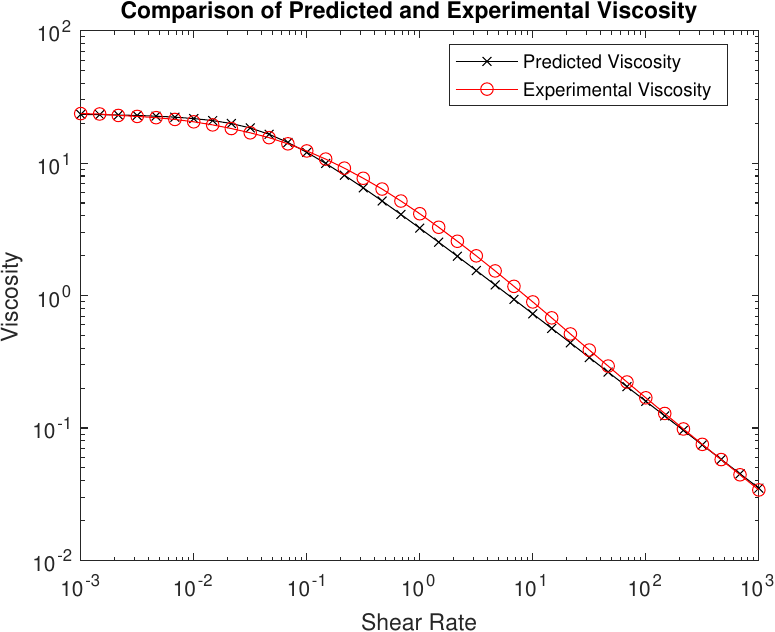}
    \caption{\label{fig:steady_shear} Comparison of experimental and fitted data showing the variation of total viscosity with respect to shear rate obtained from steady shear tests. The optimal fit data is obtained with~$\beta = 5 \times 10^{-5}$, $\lambda = 65.5,\text{s}$, and $\mu_p = 19.23,\text{Pa}\cdot\text{s}$ in the analytical expression for L-PTT model for steady shear flow.}
\end{figure}

For the PAA solution at 1\% wt. concentration, the amplitude sweep and frequency sweep resulted in storage modulus~$G^\prime = 7.2\,Pa$ and stress relaxation time~$\lambda = 6.25\, s$, respectively.
The steady shear curve (see \S\ref{sec:rheology}) is used as the basis for tuning fluid properties such as viscosity components and model parameters. To achieve this, we employ the analytical expression provided by~\cite{S_alves2001} for the steady-shear response using the L-PTT model. Figure~\ref{fig:steady_shear} presents the experimental data for total viscosity as a function of shear rate, along with the fitted curve based on the parameters $\beta = 5 \times 10^{-5}$, $\lambda = 65.5,\text{s}$, and $\mu_p = 19.23,\text{Pa}\cdot\text{s}$ derived from the L-PTT model. However, the fitted parameters deviate significantly from the experimental data.

\begin{figure}
    \includegraphics[width=0.5\columnwidth]{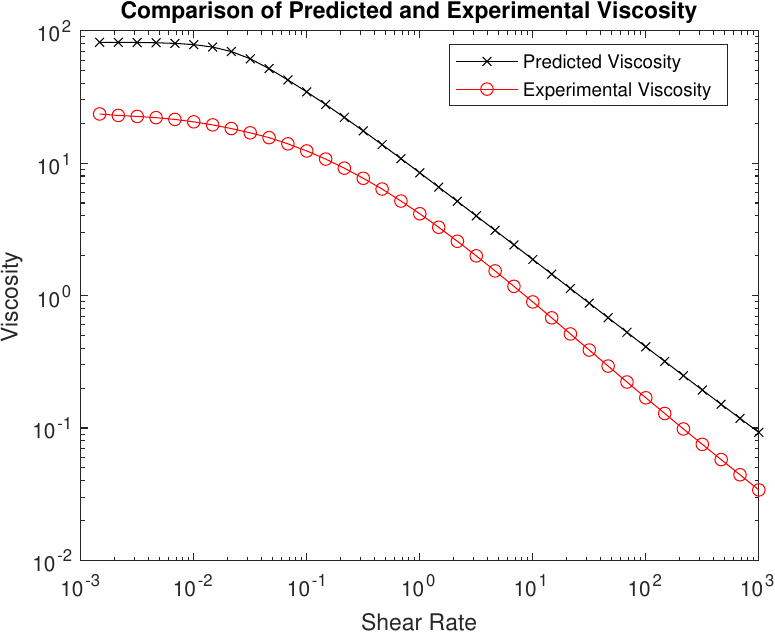}
    \caption{\label{fig:steady_shear_revise} Comparison of experimental and revised fit data for the viscosity variation in a steady shear test. The revised fit data is $\beta = 2 \times 10^{-4}$, $\mu_p = 90,\text{Pa}\cdot\text{s}$, and $\lambda = 6.25,\text{s}$ for the analytical expression of L-PTT model in a steady shear flow.}
\end{figure}

To ensure numerical stability, we adopt a slightly higher solvent-to-total viscosity ratio and set $\beta = 2 \times 10^{-4}$. Further parameter adjustments are made based on $G^\prime := \mu_p / \lambda = 7.2,\text{Pa}$, obtained from the amplitude sweep test, and the relaxation time determined from the crossover point in the frequency sweep test. The revised parameters for the numerical study are: $\beta = 2 \times 10^{-4}$, $\mu_p = 90,\text{Pa}\cdot\text{s}$, and $\lambda = 6.25,\text{s}$. The steady shear curve generated using these revised parameters is shown in figure~\ref{fig:steady_shear_revise}, alongside the experimental data. While the revised parameters still do not match the experimental data closely, they suggest a fluid with a significant elastic shear modulus~($G^\prime$) and very high relaxation times. The large relaxation time indicates that elastic energy dissipation~($\propto \lambda^{-2}$, refer~\cite{S_snoeijer2020}) is likely negligible over the time scale of the drop impact process. Thus, we proceed with these revised fluid parameters, aiming to simulate a visco-elastic material that roughly mimics the experimental data and captures the ligament formation phenomenon.

Note that different rheological models were also tried including FENE-P model~(based on finite extensibility of spring in the dumbbell model). FENE-P model with the right tuning parameters were also able to capture the ligament formation however for the numerical tests performed with different~$L_\mathrm{max}^2$, we observed the break-up of ligament before detachment from substrate and hence careful tuning of the parameters might be necessary and is not discussed in the scope of this study.

\subsection{Simulation domain and boundary conditions}
We consider a square domain measuring $8R_0$ on each side, representing only one slice of the drop impact process considering the axisymmetric flow assumption. For both liquid and gas, free-slip and no-penetration boundary conditions are applied at the domain boundaries, while a zero-gradient condition is used for pressure. To ensure that fluid can leave the domain, an outflow boundary condition is employed at the top boundary. The chosen domain size ensures that the boundaries do not influence the drop impact process. We utilize the adaptive-mesh refinement feature of Basilisk C to generate control volumes in the employed computational domain. The discretization errors in the volume of fluid tracer~($c$) and interface curvature~($\kappa$) are minimized by applying a tolerance threshold of~$10^{-3}$ and $10^{-4}$, respectively. In addition, the refinement of the grid is also performed based on the velocity components~$u_r,\,u_z$~(with a tolerance threshold, $10^{-2}$) and conformation tensor $\boldsymbol{A}$ (with a tolerance threshold, $10^{-2}$) to accurately resolve the regions with large gradients of viscous and elastic stress, respectively. We employ a grid resolution such that a minimum cell size of $\Delta = R_0/512$ is obtained, which corresponds to 512 cells across initial bubble radius.

\subsection{Initial condition}
\begin{figure}[h!]
    \includegraphics[width=0.4\textwidth]{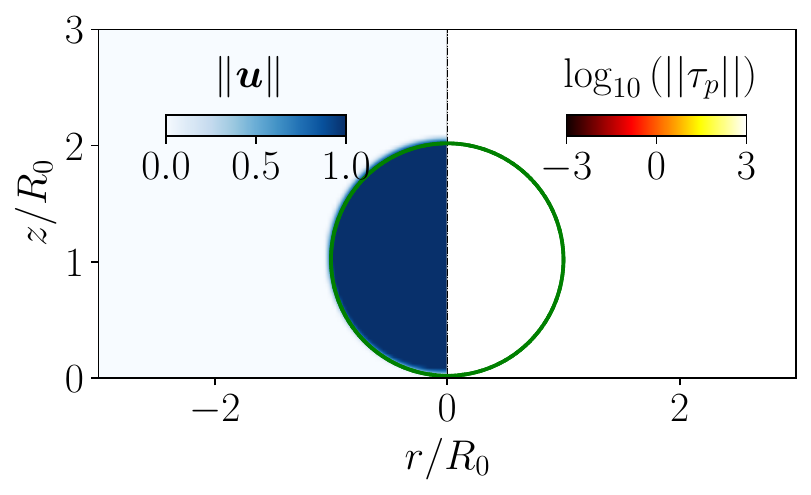}
    \caption{\label{fig:ic} Initial condition for the drop impacting on a super-hydrophobic surface. Left panel shows the non-dimensional velocity scaled with impact velocity~$U_0$ and the right panel shows the magnitude of non-dimensional polymer stress on a~$\mathrm{log}_{10}$ scale.}
    \end{figure}

The droplet is initialized very close to the wall (as shown in figure~\ref{fig:ic}) with the impact velocity of~$U_0 = \sqrt{2 g H}$, where $H$ is the release height of droplet in experiments and $g$ is the acceleration due to gravity. In this numerical investigation, we report the quantities in non-dimensional form with the length scale corresponding to~$R_0$ and the velocity scale corresponds to~$U_0$, with the pressure and stress quantities scaled by~$\rho U_0^2$. Below this point, we use the same symbols as introduced in equations~(\ref{seqn:mom})--(\ref{seqn:peqn}) to denote the non-dimensional quantities. We assume an initial stress-free condition in our study considering the computational cost to simulate the entire problem. However, if the drop motion were simulated from the release height ($H$), polymer stresses would develop as the drop approaches the substrate and may not remain zero as considered in the initial condition of this study.

\subsection{Dynamic contact angle model}
\begin{figure}
    \includegraphics[width=0.5\columnwidth]{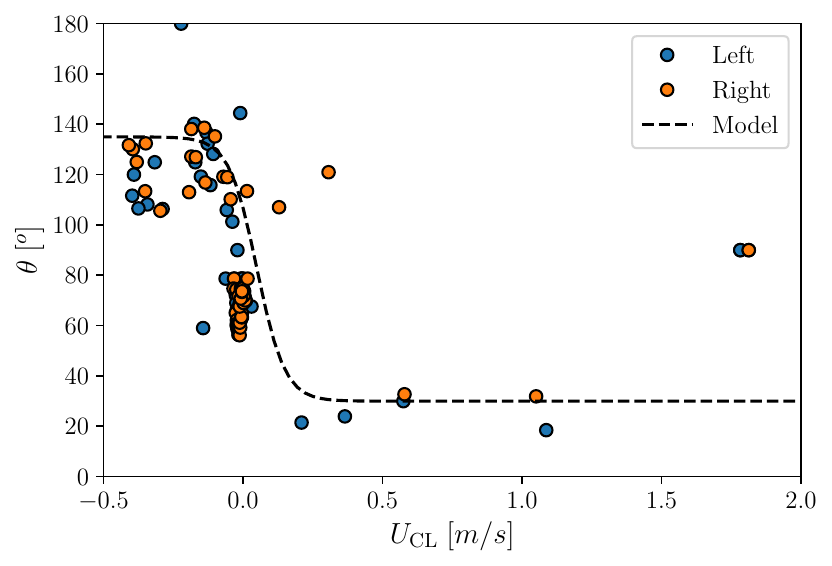}
    \caption{\label{fig:exp_vs_num} Variation of apparent contact angle with respect to the contact line speed. The experimental measurements of contact angles are indicated by markers, with \emph{Left} and \emph{Right} indicating the positions of contact points in 2D snapshots. Employed numerical model is obtained by fitting the experimental data as indicated with the dashed line.}
\end{figure}

The dynamics of the droplet on the surface is highly dependent on the formulation of dynamic contact angle, which is related to the wettability of the surface~\cite{S_yokoi2009}. In this work, we aim to simulate the viscoelastic drop impact dynamics on a super-hydrophobic surface such as Glaco, to verify if the numerical models allow for ligament formation as obtained in experiments. In particular for the numerical investigation, we have considered the case of PAA droplet with a concentration of 10,000 ppm released from a height of~$40$ cm from the substrate, although the employed numerical methodology can be suitably extended to other parameter combinations as performed in the experiments. In order to verify if the numerical simulation can reproduce the results obtained with an experiment, we employ a non-predictive dynamic contact angle model similar to~\cite{S_yokoi2009}, where the measured values of apparent contact angle obtained with the drop impact experiment is used. Figure~\ref{fig:exp_vs_num} shows the variation of contact angle with respect to the contact line speed. The imposed dynamic contact angle is of the form:
\begin{equation}
    \theta(U_\mathrm{CL}) = \frac{\pi}{180}\left(135 + 105\left(\frac{1}{1+\mathrm{exp}(-20(U_\mathrm{CL}-0.05))}\right)\right)\,,
\end{equation}

where~$U_\mathrm{CL}$ is measured at the contact point using height functions. The dynamic contact angle model simply consists in imposing a constant angle of~$135^o$ during advancing and~$30^o$ during receding phase, when inertia is dominant~(at high~Capillary number~$Ca:=(\mu_s+\mu_p)U_\mathrm{CL}/\sigma$) and an approximation for the transition between the considered angles is performed in the capillary regime~(at low~Ca).

At the contact point, there is a singularity problem as no-slip boundary condition is employed at the wall/substrate. However, in the simulation methodology the color function is advected using the face-centered velocity field which is~$\Delta/2$ away from the bottom boundary and we denote it as~\emph{numerical} slip. In addition, we have also employed a Navier slip condition during the receding phase with a Navier slip length corresponding to~$0.04R_0$. For different cases, further tuning might be required to obtain a reasonable match with the experiments.

Overall, with the above employed methodology we are able to simulate the drop impact dynamics of PAA 10,000ppm on to a super-hydrophobic substrate at high Weber numbers, as observed in the experimental study. However, it should be highlighted that the numerical methodology as employed in this study needs to be \emph{fine-tuned} if different parameter space has to be explored, as confirming to experimental observations. This implies that the present numerical methodology is not employed as a predictive tool but as a means to obtain an indication of the polymer stress distribution, provided reasonable assumption of the polymer behavior is accounted for in the employed model.

In the post-processing of numerical simulation, the wetting length~($L_w$) is measured as the distance between the axis and the contact point defined by the interface. Further, the surface pressure force ($F_p$) is calculated by integrating the dynamic pressure distribution on the substrate. Note that in order to obtain the total normal force at the substrate, the normal component contribution of viscous and polymer stress needs to be included to the surface pressure force.

\end{document}